# Cometary Observations in Light-Polluted Environments: A case study of Interstellar Comet 2I/Borisov

Josep M. Trigo-Rodríguez[1,2]
Damya Souami[3,4],
Maria Gritsevich[5,6,7],
Marcin Wesołowski[8], and Gennady Borisov[9]

[1] Institute of Space Science (ICE-CSIC), Campus UAB, Carrer de Can Magrans, s/n, Barcelona, Catalonia, Spain. E-mail: trigo@ice.csic.es
[2]Institut d'Estudis Espacials de Catalunya (IEEC), Esteve Terradas 1, Edifici RDIT, Ofic. 212, Parc Mediterrani de la Tecnologia (PMT), Campus del Baix Llobregat - UPC, 08860, Castelldefels, Barcelona, Catalonia, Spain,
[3] LIRA, CNRS, Observatoire de Paris, Université PSL, Sorbonne Université, Université Paris Cité, CY Cergy Paris Université, 92190 Meudon, France
[4] naXys, University of Namur, 11 rue de Bruxelles, Namur 5000, Belgium
[5]Swedish Institute of Space Physics (IRF), Bengt Hultqvists väg 1, 981 92 Kiruna, Sweden
[6]Faculty of Science, University of Helsinki, Gustaf Hallströmin katu 2, FI-00014 Helsinki, Finland
[7]Institute of Physics and Technology, Ural Federal University, Mira str. 19, 620002 Ekaterinburg
[8] University of Rzeszów, Faculty of Exact and Technical Sciences, Institute of Physics, Pigonia 1 Street, 35-310 Rzeszów, Poland.
[9]MARGO astronomical observatory, Nauchnij, Crimea.

**Abstract:** Comets and asteroids have long captured human curiosity, and until recently, all documented examples belonged to our Solar System. That changed with the discovery of the first known interstellar object, 1I/2017 U1 ('Oumuamua), in 2017. Two years later, on August 30, 2019, Gennady Borisov discovered a second interstellar object, 2019 Q4, which was officially designated 2I/Borisov. From its initial images, the object's diffuse appearance hinted at its cometary nature. To better understand the photometric evolution of comet 2I/Borisov as it traveled through the inner Solar System, we compiled observations using medium-sized telescopes. This data is crucial for gaining insights into its size and composition, as well as how such objects, after millions of years in interstellar space, behave when exposed to the Sun's radiation. Given that 2I/Borisov is the first interstellar comet ever observed, constraining its behavior is of great scientific interest. In this paper, we present photometric data gathered from observatories in Crimea and Catalonia, highlighting the importance of systematic photometric studies of interstellar objects using meter-class telescopes. Our observations showed a steady increase in the comet's brightness as it approached perihelion, likely due to the slow sublimation of ices. Over the five-month pre-perihelion observation period, we did not detect any significant changes in magnitude. The analysis of observations reveals a steady increase in comet 2I/Borisov brightness as it approached perihelion, likely due to the sublimation of ices, with no observable outbursts during the five-month pre-perihelion period. Additionally, we discuss the challenges in ground-based observation of comets posed by light pollution today, particularly in urban areas, where visual observations are severely limited. Using sample surface brightness measurements, we demonstrate the impact of light pollution and outline the importance of systematic photometric studies for interstellar objects.

1. Introduction

The existence of interstellar objects of asteroidal or cometary nature was predicted long ago (see e.g. Morbidelli, 2008), but clear evidence was not obtained until 2017 with the detection of 1I/'Oumuamua (Peña-Asensio et al., 2024). A careful study of the imagery of this visitor revealed that it was an extremely oblong shape body, and no directly observable signs of coma formation or jet activity was reported (Meech et al., 2017). It is also remarkable that the albedo of Oumuamua is not yet well constrained. We could think that such a type of interstellar visitor should be dark, but a high-albedo scenario is not fully ruled out (see Trilling et al. 2018). Despite of this apparent lack of cometary activity in 1I/'Oumuamua some studies suggest, from an accurate analysis of astrometric data, a non-gravitational acceleration compatible with a comet-like outgassing (Micheli et al., 2018). Such a scenario is still debated, other authors suggest that the body was only driven by radiation pressure. Obviously, there is a significant debate of the real nature of these interstellar visitors because their accretion in distant protoplanetary disks or nebula should produce bodies with distinctive chemical abundances (Seligman & Laughlin, 2020) unable to survive long stays in the interstellar medium (ISM) as proposed by Hoang and Loeb (2020). In any case, deciphering the real nature and origin of these bodies requires all our observational efforts during the short periods of time in which they become observable from Earth (Hallat and Wiegert, 2020; Luu et al., 2020).

The relevance of these studies is out of doubt because we all want to be protected from unexpected catastrophic encounters (Mainzer et al., 2021; Trigo-Rodríguez, 2022). To avoid them we need to discover the minor bodies that pose a risk to us well in advance so that palliative methods can be arranged. In fact, DART NASA mission has started a new era of active planetary defense (see e.g., Daly et al., 2023). Having success depends on our ability to predict the consequences of impacts (Gritsevich et al. 2012, 2013; Senel et al., 2023) and detect flagged minor bodies with sufficient time to implement deflections techniques (Clemens et al., 2020, Chen et al., 2023). At the top of the list of dangerous celestial visitors are fast-moving objects, originating from highly eccentric or even hyperbolic orbits. Since kinetic energy is proportional to velocity squared, these high-velocity objects carry significantly more energy. Moreover, their high velocities leave less time for detection and mitigation, making early identification crucial for planetary defense efforts.

The study of interstellar comets can also contribute to increase our understanding of the dynamical evolution of comets in the early stages of protoplanetary disks, including our own (Morbidelli, 2008; Weissman et al., 2020). We know that the outer solar system planet distribution changed by the migration of the giant planets that produced the subsequent scattering of planetesimals (Walsh et al., 2011). Obviously, such a process caused deep changes in the distribution of small bodies in our planetary system, and it has potential to explain a significant escape of bodies in interstellar orbits (Walsh and Morbidelli, 2009). It is reasonable to think that such early processes causing strong orbital resonances occurred as well in other planetary systems. As a result, it is theorized that a significant part of the population of bodies crossing the orbits of giant planets was scattered inwards and outwards to build the currently known distribution of minor bodies populations, but other parts were probably lost in interstellar space (Morbidelli, 2008). The consequence of such processes are the bodies finally stored into different cometary reservoirs e.g. the Scattered Disk of the Kuiper Belt (KB). Another reservoir of long period comets is the Jupiter Family Comet (JFC) sample that probably originated from the KB region thought a Neptune-based transfer mechanism (Morbidelli, 2008).

Follow-up and discovery surveys addressed to catalogue all minor bodies in our solar system were thought to be effective in making real progress in our understanding of the observational study of small solar system objects. These automatic programmes were thought to leave little chance for the discovery and follow-up by amateurs, but this seems to be wrong. On August 30$^{th}$, 2019 Gennady Borisov discovered the first interstellar comet using a 0.65 m telescope, for which the provisional object 2019 Q4 received the official designation of 2I/Borisov. This significant achievement, obtained when this comet was fainter than +18 magnitude, was made by an amateur astronomer. Such an amazing discovery is emphasising that semi-professional observatories still have something to say in front of the search programmes made nowadays by professional automatic surveys. From time to time a triumph of small-telescopes dedicated astronomy in the field of solar system minor bodies is required to motivate amateurs to contribute actively into scientific research. Over the years, our research team has participated in the systematic photometric follow-up of interesting comets that experience unexpected photometric changes as outburst (Trigo-Rodríguez et al., 2008, 2010; Gritsevich et al., 2022, 2025ab), therefore we decided to perform a dedicated follow-up of this fascinating object as well.

Astronomical research depends on access to dark skies, and excessive artificial lighting threatens the viability of observatories, particularly those near urban centers. Many professional observatories invest in costly mitigation strategies, such as adaptive optics, to compensate for degraded sky quality. Moreover, large-scale sky surveys and deep-sky imaging projects require long exposure times, which become less effective due to artificial sky brightness. The economic impact extends beyond wasted electricity. It also affects the efficiency of astronomical research, potentially delaying discoveries and technological advancements.

Consequently, the phenomenon of light pollution is a fundamental concern for visual astronomical observations, particularly for celestial bodies of relatively low brightness like e.g. comets are. The causes of this phenomenon can be categorized into natural and artificial factors. Natural sources include the Moon, planets, comets, and stars, while artificial sources stem from human activity, notably urbanization and the expansion of artificial lighting.

Light pollution affects nearly every country, especially developed and rapidly developing nations. In Europe, central European countries like e.g. Poland are prime examples, with virtually no naturally dark areas free from light pollution. The only exceptions are two designated dark sky parks (Bieszczady Dark Sky Park and Izerski Dark Sky Park) and regional initiatives aimed at preserving natural darkness. Over the South of Europe some locations, like the ones here used to perform the observations, remain relatively far from the excessively illuminated coast, and still have reasonable sky quality (Table 1). Given these challenges, systematic local monitoring of night sky brightness is crucial for astronomical observations. Light pollution reduces the contrast between celestial objects and the night sky, significantly hindering the visibility of faint objects, particularly near large cities. Even at considerable distances from urban centers, residual light pollution can render observations of faint celestial bodies, such as comets, nearly impossible (Wesołowski 2023).

Beyond astronomy, light pollution negatively impacts human health, safety, ecology, the economy, and aesthetics, making it a significant environmental issue. The consequences of this phenomenon have been systematically studied for over 30 years, particularly by astronomers (Aube & Kocifaj 2012; Bara 2014; Biggs et al. 2012; Cavazzani et al. 2015; Cinzano et al. 2001a, 2001b; Cinzano & Elvidge 2004; Falchi et al. 2023; Garstang 1989; Gronkowski et al. 2018; Hanel et al. 2018; Holker et al. 2021; Kocifaj et al. 2023; Kaushik et al. 2022; Kyba et al. 2017, 2023; Sanchez et al. 2017, 2021, 2022; Wesołowski 2019, 2020, 2023).

To measure artificial light pollution in the night sky, the Sky Quality Meter (SQM or SQM-L) is commonly used. This simple photometer quantifies radiance, the amount of light emitted from a specific area of the sky, converting it into surface brightness units (mag/arcsec²). The SQM-L model includes a built-in lens that narrows the measurement range to 20° (compared to 84° in the standard SQM model), improving accuracy by minimizing interference from horizon light pollution. According to technical documentation, the device has a measurement accuracy of ±10% of the recorded value.

It is important to emphasize that 2I/Borisov significant discovery, obtained when this comet was fainter than +18 magnitude, was made by an amateur using a mid-size telescope (Borisov et al., 2019; Masi et al., 2019). The comet, distanced 2.98 AU from the Sun (Guzik et al., 2019; Opitom et al, 2019), had a clear hyperbolic orbit, with an eccentricity >3. Recent spectropolarimetric observations of this comet revealed that the polarization of 2I/Borisov is higher than what is typically measured for Solar System comets (Bagnulo et al., 2021). The spectroscopic observations found that the comet has a featureless spectrum (de Leon et al., 2019). The comet's nucleus radius was first estimated to be between 0.7 km and 3.3 km by dedicated observations (Fitzsimmons et al., 2019). On the other hand, Jewitt et al. (2020) constrained the nucleus size to $r \leq 0.5$ km through Hubble Space Telescope (HST) observations, using a geometric albedo of 0.04, and measuring the nongravitational effect (non-detection), which give consistent results. The nongravitational effect was later on better constrained by Hui et al. (2020), confirming the nucleus' size. Future early discovery of these interstellar objects will give us the chance to intercept them to constrain these parameters and test the concept of the ESA Comet Interceptor mission, planned to be the first spacecraft to visit an interstellar object during its short transit through the inner Solar System (Snodgrass and Jones, 2019; Hasinger, 2020; Jones et al., 2024).

The main goal of this paper is to present a photometric study of comet 2I/Borisov, and exemplifying the relevance of systematic photometric follow-ups using mid-sized telescopes to infer the level of dust activity. We were particularly motivated by the opportunity to study for the very first time the photometric evolution of an interstellar comet during its transit through the inner solar system. Perhaps the photometric behaviour of this body could give us clues on the processing of volatiles after a long journey through the interstellar medium, but also to get insight on other physical processes (Yang et al., 2021), all of which needed to better constrain its size and nature. In this paper, we compile the photometric measurements made from Catalonia and Crimea using meter-class instruments.

## 2. Observability determination

Determining the visibility of a comet through a telescope is a complex task influenced by several factors, including the comet's actual brightness and the conditions of the observing environment. One key challenge **is** artificial and natural light pollution, which can significantly hinder comet visibility. For instance, in urban areas, sky pollution makes visual observation of comets nearly impossible. Studies conducted in Rzeszów, southeastern Poland, show that the city's light pollution extends up to 25 km, limiting opportunities for clear sky observations (Wesołowski, 2019, 2023). In such conditions, only very bright comets, or those passing close to Earth, such as 62P/Tsuchinshan, 144P/Kushida, C/2021 S3 PanSTARRS, 12P/Pons–Brooks and C/2023 A3 Tsuchinshan–ATLAS can be observed.

To improve observability, telescopes are widely used, yet astronomers have debated since the 1960s about the faintest objects that can be detected by a given telescope. The limiting brightness, or magnitude, of a comet visible through a telescope can be calculated using the following relationship:

$m = n_i + 5 \cdot log(D_{ob})$,  [1]

where $m$ is the limiting magnitude of the comet that is visible through a telescope with an objective diameter of $D_{ob}$ expressed in millimeters, and $n_i$ is a normalization constant. The normalization constant $n_i$ is influenced by multiple parameters, including the surface brightness of the night sky, the magnification of the telescope, the observer's eye pupil diameter, and individual visual acuity. It also depends on atmospheric conditions such as airglow and light pollution, which can significantly impact the visibility of faint objects. Based on conducted studies, $n_i$ has been found to range from 1 to 12 (Schaefer, 1990). The distribution of the comet limiting magnitude as a function of the telescope's objective diameter is shown in Fig. 1.

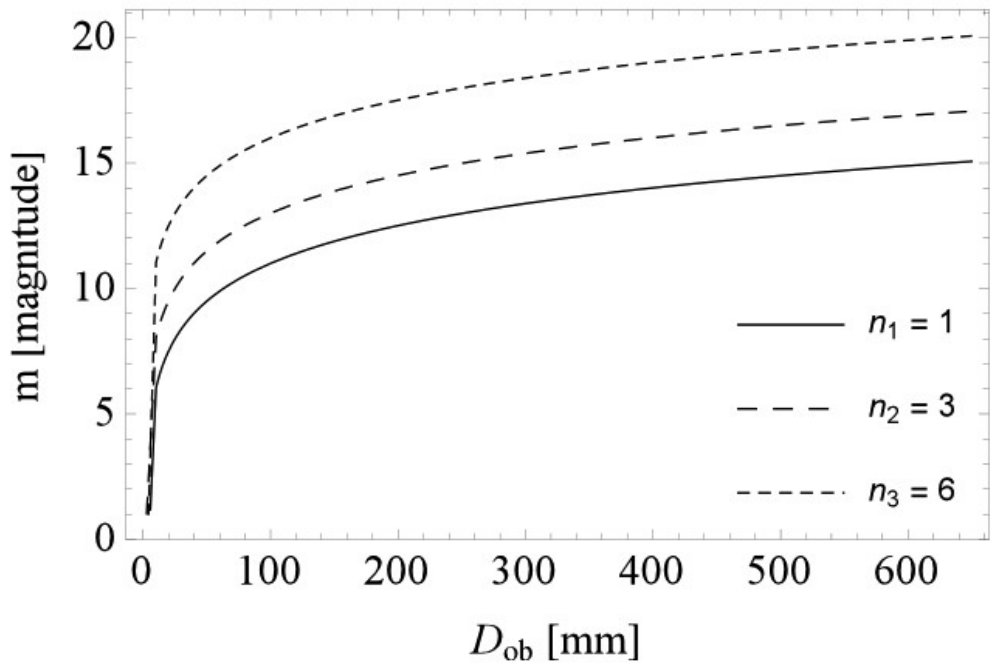

Figure 1. Limiting magnitude as a function of the objective diameter ($D_{ob}$)

As demonstrated in Fig. 1, the distribution shows that as the diameter of the telescope's objective increases, the limiting magnitude increases as well. This means that larger telescopes have a greater range and can detect fainter comets. For a telescope with a 650 mm objective, depending on the normalization constant, the limiting brightness is found to range between +15.1 and +20.1 magnitudes, making it capable of observing much fainter celestial bodies compared to smaller telescopes. The main limitation during observation is the surface brightness of the night sky and its effect on the visibility of the comet through the telescope.

3. Observations and data reduction.

The three observatories involved in this study are listed in Table 1, all being *Minor Planet Center (MPC)* contributors. From Catalonia we observed comet 2I/Borisov using *Telescopi Joan Oró (TJO, MPC code C65)*, a 1 m-class telescope working in a completely unattended manner. It operates in the *Observatori Astronòmic del Montsec (OAdM: www.oadm.cat)*, a site devoted to host astronomical research facilities under dark skies. TJO CCD detector was an iKon XL camera, with a back-illuminated 4k×4k chip manufactured by Andor. This CCD camera provides a FoV of 30×30 arcmin$^2$, with a resolution of 0.4 arcsec for a pixel size of 15 μm. The much smaller 0.25 m telescope at Montseny Observatory (MPC B06) was using a ST8 camera, and was able to image the comet working with a resolution of 1 arcsec/pix (see Table 2).

From Crimea the comet was monitored from *MARGO observatory* (MPC Code L51) where one of the authors, Gennady Borisov, used a 0.65 m in diameter Hamilton telescope at F/1.52 with a focal length of 991 mm. Such system with a FLI ML16803 CCD provided a FOV of 128×128 arcmin$^2$, and a resolution of 1.87 arcsec/pix, operated without filters.

Table 1. Observatories and instruments used in this study. As a first approach we indicate the average sky quality meter (SQM) obtained from: https://www.lightpollutionmap.info/ . Accurate C65 sky quality measurements are available thanks to the TESS network at: https://tess.dashboards.stars4all.eu/d/datasheet_stars831/stars831?orgId=1

| Observatory | MPC code | Telescope | SQM (mag./arc sec$^2$) | Bortle class |
|---|---|---|---|---|
| MARGO Astron. Obs., Crimea | L51 | SC 65 **F/1.52** | 21.80 | 2/3 |
| TJO, Montsec (OAdM), Cat | C65 | RC 80 F/9.6 | 21.77 | 2 |
| Montseny, Cat | B06 | SC 25 F/10 | 20.27 | 4-5 |

Given the faintness of this comet, we have decided to make unfiltered CCD imaging, except for few R-filtered images obtained from C65 (Table 2). We have used similar reduction procedures than in the previous papers (Trigo-Rodríguez et al., 2008; 2010), but this time the photometry was measured for a circular aperture of about 20 arcsec centered at the comet's false nucleus (Stetson, 2000). Using these standards, we were able to quantify the typical data accuracy to be better than 0.1 mag.

Concerning the photometric measurements, to perform the photometric reduction from the comet images, we used a photometric aperture of 10 arcsec surrounding the false nucleus. The data compilation shows that the comet's intrinsic brightness increased monotonically following a predictable pattern (see Table 2). The data reduction basically consisted of bias subtraction, flat-field correction and flux calibration. The photometric reduction procedure for small-sized telescopes was performed by using an additional software package called LAIA (after the Laboratory for Astronomical Image Analysis), successfully tested for obtaining high-precision stellar photometry (García-Melendo & Clement 1997; Escolà-Sirisi et al. 2005). Using these standards, we were able to quantify the typical data accuracy to be better than 0.05 mag. Then, our given R* is the apparent unfiltered comet magnitude (not using a R Johnson Cousin filter, except for C65 data).

Table 2. Observational data: Julian date, comet unfiltered R* magnitude (except for C65 data, obtained using a R Johnson Cousin filter), heliocentric distance (r), geocentric distance (Δ), and A(θ)fρ values.

| Date | Julian Date | R* Mag | r (AU) | Δ (AU) | FOV (arcmin$^2$) | A(0)fρ (cm) | MPC code |
|---|---|---|---|---|---|---|---|
| 29 Aug. 2019 | 2458725.5 | 18.0 | 2.98626 | 3.72139 | 128×128 | 90±16 | L51 |
| 30 Aug. 2019 | 2458726.5 | 17.8 | 2.97030 | 3.69919 | 128×128 | 96±18 | L51 |
| 31 Aug. 2019 | 2458727.5 | 18.0 | 2.95441 | 3.67698 | 128×128 | 93±17 | L51 |
| 1 Sep. 2019 | 2458728.5 | 18.1 | 2.93858 | 3.65476 | 128×128 | 95±18 | L51 |
| 2 Sep. 2019 | 2458729.5 | 17.8 | 2.92282 | 3.63254 | 128×128 | 97±19 | L51 |
| 3 Sep. 2019 | 2458730.5 | 17.9 | 2.90713 | 3.61032 | 128×128 | 100±20 | L51 |
| 4 Sep. 2019 | 2458731.5 | 17.9 | 2.89155 | 3.58811 | 128×128 | 100±19 | L51 |
| 6 Sep. 2019 | 2458733.5 | 17.7 | 2.84508 | 3.52150 | 128×128 | 103±21 | L51 |
| 8 Sep. 2019 | 2458735.5 | 17.6 | 2.82976 | 3.49932 | 128×128 | 105±22 | L51 |
| 9 Sep. 2019 | 2458736.5 | 17.5 | 2.81455 | 3.47717 | 128×128 | 110±25 | L51 |
| 12 Sep. 2019 | 2458739.5 | 17.6 | 2.76892 | 3.41072 | 30×30 | 104±32 | C65 |
| 18 Sep. 2019 | 2458744.7 | 17.4 | 2.67045 | 3.25915 | 30×30 | 125±19 | C65 |
| 28 Sep 2019 | 2458755.5 | 17.3 | 2.54157 | 3.06334 | 128×128 | 128±27 | L51 |
| 1 Oct. 2019 | 2458758.5 | 17.2 | 2.50183 | 3.00001 | 128×128 | 135±29 | L51 |
| 3 Oct. 2019 | 2458760.5 | 17.3 | 2.47593 | 2.95824 | 128×128 | 130±26 | L51 |
| 9 Oct. 2019 | 2458766.5 | 17.1 | 2.40122 | 2.83543 | 128×128 | 130±25 | L51 |
| 25 Oct. 2019 | 2458782.5 | 16.9 | 2.22726 | 2.53166 | 128×128 | 125±21 | L51 |
| 26 Oct. 2019 | 2458783.5 | 16.9 | 2.21777 | 2.51407 | 128×128 | 123±20 | L51 |
| 1 Nov. 2019 | 2458789.5 | 16.8 | 2.16461 | 2.41270 | 128×128 | 128±23 | L51 |
| 2 Nov. 2019 | 2458790.5 | 16.9 | 2.15640 | 2.39655 | 128×128 | 126±22 | L51 |
| 8 Nov. 2019 | 2458796.6 | 17.0 | 2.11066 | 2.30296 | 128×128 | 108±18 | L51 |
| 6 Dec. 2019 | 2458824.6 | 16.8 | 2.00680 | 2.00320 | 128×128 | 110±23 | L51 |
| 7 Dec. 2019 | 2458825.5 | 16.9 | 1.99860 | 1.99250 | 35×28 | 105±21 | B06 |

To characterize the activity of dust in the coma, we used the photometric data to infer the Afρ values using a similar approach as in our previous work (Trigo-Rodríguez et al., 2010, Gritsevich et al., 2025b). This time we took into consideration that Afρ can be measured in terms of magnitudes as demonstrated in (Gillan, 2025, page 16):

$$Af\rho = \frac{4r_h^2 \Delta^2}{\rho} \cdot 10^{0.4(m_\odot - m)} \qquad [2]$$

Where $r_h$ is the heliocentric distance in astronomical units (au), and $\Delta$ is the geocentric distance measured in cm. On the other hand, $m_\odot$ and $m$ are the apparent magnitudes of the Sun and comet respectively. $\rho$ is the photometric aperture, chosen here as 10 arcsec. For the absolute magnitude of the Sun we used the value of -26.97 for the R Cousins band given by Willmer (2018). To obtain the Af$\rho$ (corrected at zero phase angle) we used Schleicher (2010) composite dust phase function, having into account that the phase angle changed in the observational period between 20 and 28º. The errors given in the Af$\rho$ values indicated in the Table 2 were obtained using the method of partial derivatives of [2].

4. Discussion.

Concerning the observability period of 2I/Borisov, it was visible from the northern hemisphere between August and November, 2019. During this observational period the heliocentric distance ranged from 2.986 to 1.999 AU at the time of our observations (Table 2). The comet exhibited a blurry coma is exemplified in TJO image shown in Fig. 2. The comet's overall appearance, also exhibiting one arcmin diffuse tail is seen in the selection of images obtained from Crimea (see Fig. 3). The object crossed the ecliptic plane on October 26th, and the celestial equator on Nov. 13th, 2019, entering the southern sky. On December 8th, 2019, the comet reached perihelion and became too low from our observatories, so our photometric data is restricted to the pre-perihelion approach. In general, our observations using medium-size telescopes identified 2I/Borisov with a quite diffuse coma. Despite of our relatively low spatial resolution, our results support an asymmetry in the dust coma that has been explained as produced by a thermal lag on the rotating nucleus (Kim et al., 2020). In general, our CCD observations noticed little changes in comet 2I/Borisov coma during the months in which we performed the photometric follow-up. This result is revealing very little differences with the observations made with larger instruments (Guzik et al., 2020; Kim et al., 2020), thus demonstrating the potential of mid-sized instruments to discover, track, and perform photometric follow-up of these elusive objects.

The overall weak and tenuous appearance of 2I/ Borisov also provides significant clues about the need of performing the searches of interstellar visitors using mid-size space telescopes. In the case of the three observatories involved in this work (Table 1) we can realize that the sky quality of C65 and L51 are similar, roughly among Bortle sky 2, while the small instrument at B06 is strongly affected by light pollution with a Bortle class 5. In the case of a diffuse object as 2I/ Borisov, we measured typical signal/noise (S/N) ratios of about 1.20 for L51 and C65, while it was S/N<1.1 for B06, close to the limit but noticeable in stacked images. Obviously, it is particularly challenging to get accurate astrometric data for diffuse objects with such small S/N ratios.

Consequently, our data exemplifies why we think that unbiased observations from space are needed to increase the rate of discoveries of these elusive objects. A solution can be to set dedicated telescopes in space to patrol the sky and being able to discover and track interstellar objects, in principle compatible with the searches of potentially hazardous asteroids (PHAs). 2I/Borisov's orbital geometry suggests that, even when interstellar comets could probably survive in the interstellar environment over timescales of tens to hundreds of Myrs, some of these interstellar visitors could be younger (Bailer-Jones et al., 2019; Hallatt and Wiegert, 2020). In these fortuitous encounters they can experience a significant outgassing activity when approaching stars like our Sun. In any case, most of these comets could experience a quiescent behaviour, probably due to a subtle sublimation of ices, and the peculiar inherited diffuse properties (Bagnulo et al., 2021). In fact, Kim et al. (2020) found that the depth of the ice layer sublimating from the comet was comparable to, or smaller than, the meter-thick layer expected to be processed by cosmic rays. In consequence, the level of outgassing and dust production expected for interstellar bodies could be limited because the outer layer has been processed over millions of years, although it could preserve beneath a quite pristine interior. A *Comet Interceptor* type mission with a drill instrument would get into the primordial layers of the comet (Snodgrass & Jones, 2019). We should consider the possibility that a moderate cometary activity for these visitors could be due to a significant fraction of the comet surface being covered by a refractory layer produced by cosmic ray irradiation produced during long exposure to interstellar space (Cooper et al., 2003; Kim et al., 2020).

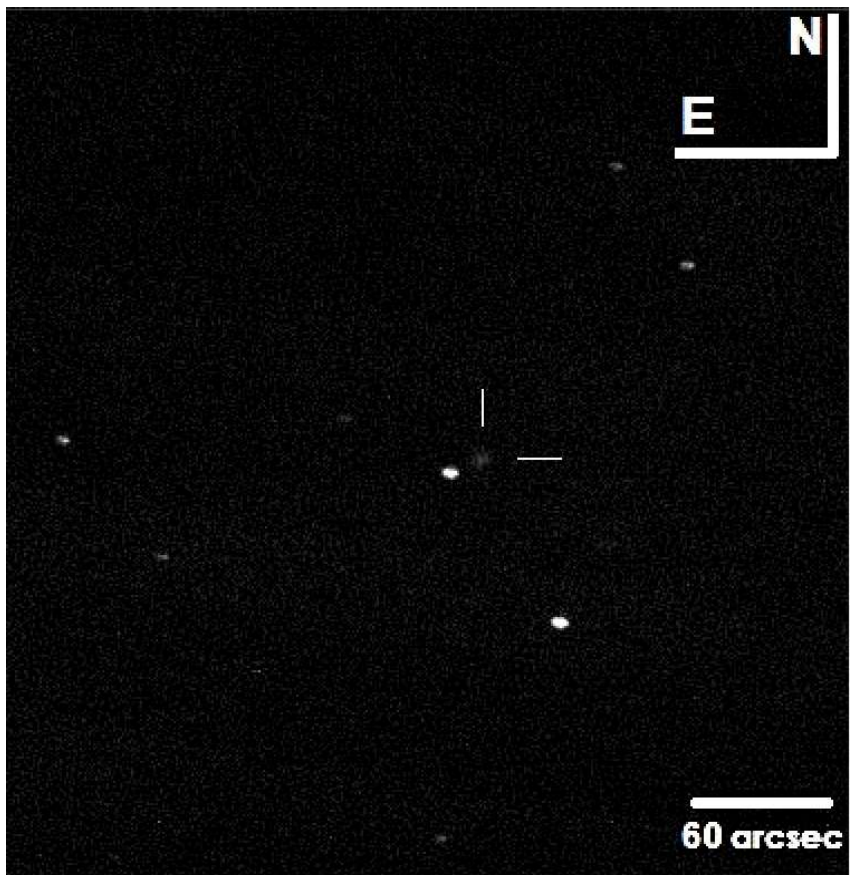

Figure 2 The diffuse appearance of the coma of 2I/Borisov in a 5 minutes' exposure TJO image obtained on September 18th, 2019 (JD: 2458744.7)

In general, small photometric changes are found in Fig. 3 as 0.1-0.2 magnitude changes are noticeable between successive nights' images, but they can be attributed to the effect of changing atmospheric extinction and stellar fields. We therefore conclude that there is no convincing photometric variability detected within each night, neither along the observation period, at least to the precision levels reached by our photometric sequence. Nevertheless, given the aperture of 10 arcsec surrounding the false nucleus, it is likely that some photometric variations may have been systematically diluted by the photometric signal originated by the coma. Equation 1 provides a 3$^{rd}$ order polynomial fit to the measured photometric points (*x* represents the Julian Date):

$$R_{mag} = -10^{-6} x^3 + 8.0468 x^2 - 2\times10^7 x + 2\times10^{13} \qquad [2]$$

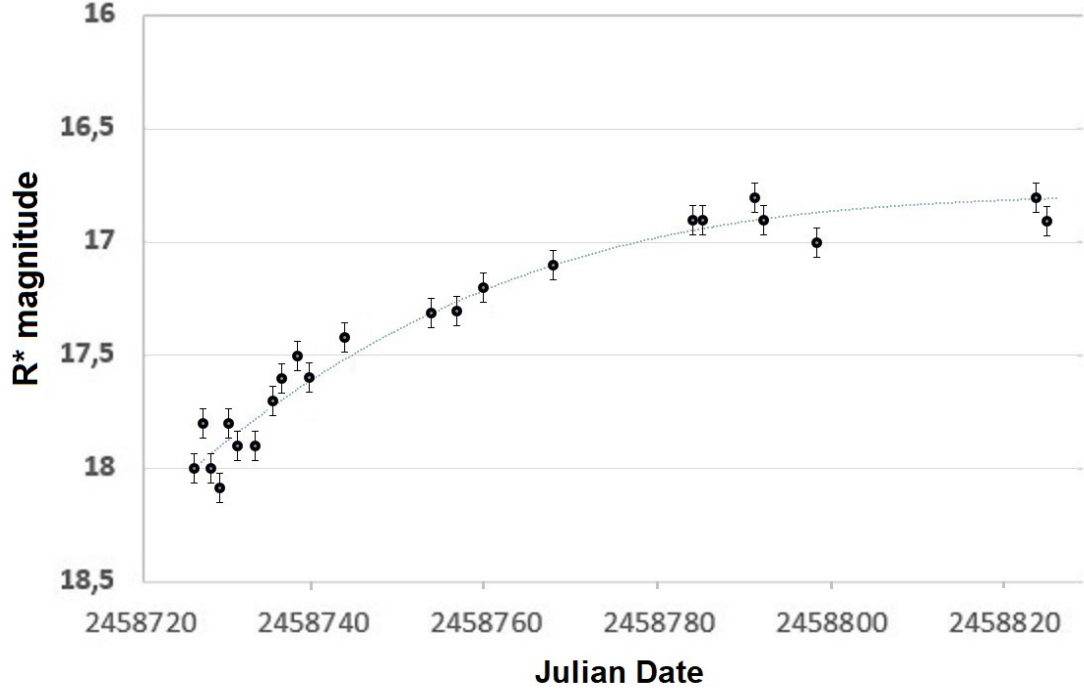

Figure 3. Unfiltered, close to "R" magnitude for comet 2I/Borisov as a function of the time in Julian Date. Photometric accuracy is about 0.1 magnitudes.

On the other hand, the dust production can provide additional clues on the behaviour of interstellar objects. Our observations started with the comet discovery at a heliocentric distance of 2.986 AU, and at a geocentric distance of 3.721 AU. At that moment the measured Afρ activity value was between 90-100 cm (see Table 2). Over the following months the Afρ values increased to 135 cm (Table 2), with a slightly increase in early November when the comet was at 2.164 AU from the Sun and still approaching Earth: 2.412 AU. It is certainly difficult to estimate the dust production rate using Afρ because it depends of several model assumptions. We have used for the computation a radius of 10$^4$ km around the nucleus and the model parameters proposed by Fitzsimmons et al. (2019): single dust grain diameter of 1 μm, grain albedo of 0.04, density of 1000 kg m$^{-3}$, and radially outflowing dust velocity V$_d$=100 m s$^{-1}$, with Q(dust)≅1 kg s$^{-1}$. Our Afρ values (Table 2) are slightly lower but, having into account the different telescope size and spatial resolution, they seem comparable to the ones obtained by Fitzsimmons et al. (2019) for Sept. 20, 2019: Afρ =143 ± 10 cm in the R band."

It is important to remark that our observations are consistent with a decline trend noticed in the intrinsic brightness of the comet as described by other authors (Hui et al., 2020, Kim et al., 2021). In fact, our data reveals that during its approach to Earth during November and December, dust production slightly decreased. On the other hand, the inferred Afρ values are also consistent with the ones obtained by Clements (2021). Such behaviour is exemplified with the compilation of images in Figure 4.

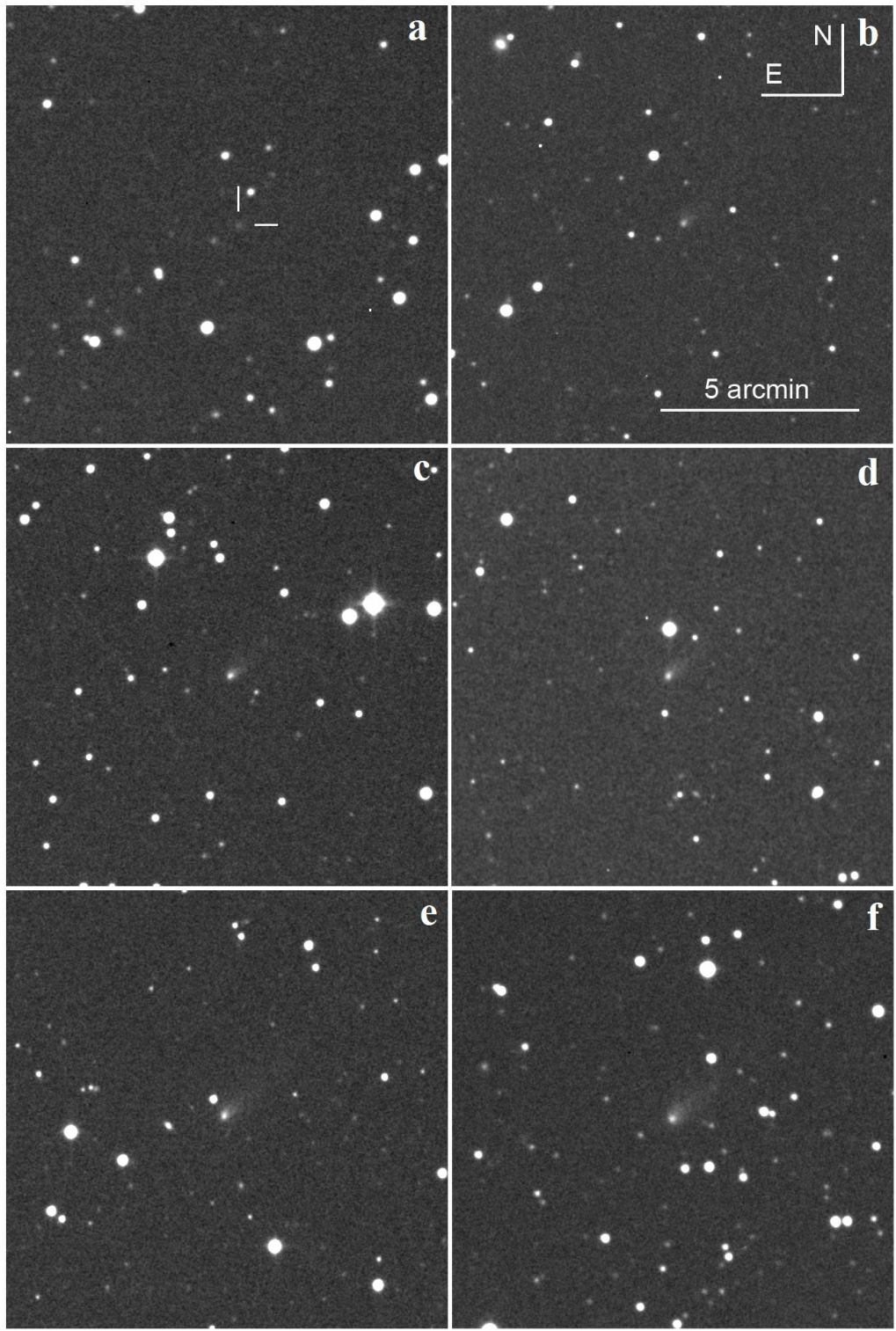

Figure 4. Images of 2I/ Borisov, centered in a FOV of 10 arcmin as recorded from MARGO observatory in Crimea. In first image the comet location is marked, while in the other images exhibits a clear coma. Field orientation is common and shown in b. a) Aug. 31, b) Sept. 28, c) Oct. 9, d) Oct. 28, e) Nov. 8, and f) Dec. 6, 2019.

The homogeneous coma appearance seems to be consistent with observations made from much larger instruments like the Gran Telescopio Canarias (GTC), the Gemini North Telescope in Hawai'i, or the Hubble Space Telescope (HST) (de León et al., 2019; Guzik et al., 2020; Jewitt et al., 2020). The visible spectrum of this comet was obtained using GTC, being similar to other Oort cloud cometary nuclei, matches quite well the D-type class defined by DeMeo et al. (2009). Jewitt et al. (2020b) concluded from the analysis of high-resolution HST images that the coma brightness can be explained if it is produced by scattering of solar light by particles with a characteristic size ~0.1 mm ejected anisotropically. To reach such a conclusion they completed a detailed convolution modelling of the coma surface brightness that allowed them to constrain a spherical-equivalent nucleus radius r≤ 0.5 km assuming a typical geometric albedo of 0.04 (Jewitt et al., 2020). In addition, interferometric observations of this comet with the Atacama Large Millimeter/submillimeter Array (ALMA) have been performed by Yang et al. (2021), finding evidence for some degree of compaction in the dust particles, which are mm sized according to their observations.

Concerning the overall structure of the comet, HST observations have revealed that the comet's nucleus could have disrupted into two pieces in early March, 2020, perhaps pointing towards being a fragile and inhomogeneous core (Jewitt et al., 2020b). In addition, it seems that small light-curve variations in 2I/ Borisov could be hidden by the dust-dominated light-scattering cross-section observed using the HST (Bolin and Lisse, 2020).

Observations of 2I/Borisov support the hypothesis that its nucleus was most likely sub-km sized (Hui et al. 2020; Jewitt et al. 2020, 2020b, 2020c; Kim et al., 2020). If they are correct, it is clear that this interstellar comet brought good news for astronomy: km-sized interstellar comets can be also discovered and tracked using mid-sized instruments, with the intrinsic limitations imposed by the telescope diameters. New comet discoveries will increase our statistics, particularly useful to quantify the frequency in which these bodies strike Earth. According to Jewitt et al. (2020), a 100 m size scale interstellar body should strike Earth every one to two hundred million years.

5. Conclusions.

From an astronomical perspective, the implementation of shielded lighting is crucial for reducing skyglow, which severely interferes with ground-based observations. Light pollution introduces artificial brightness that lowers the contrast between celestial objects and the night sky, making it difficult to observe faint astronomical bodies, such as comets and their evolving dust trails. These features are often only visible in the aftermath of very bright outburst events (Trigo-Rodríguez et al., 2008b; Lyytinen et al., 2013; Nissinen et al., 2021; Ryske et al., 2022; Gritsevich et al., 2022, 2025a), but excessive artificial illumination severely limits their detectability (Wesołowski, 2023). In particular, the use of LED lights with warmer color temperatures helps minimize blue light emissions, which scatter more efficiently in the atmosphere and significantly contribute to sky brightness. Reducing blue light is particularly important for preserving the visibility of deep-sky objects and maintaining the natural conditions necessary for precise astronomical measurements.

Our photometric study of 2I/Borisov during its pre-perihelion approach revealed a uniform behaviour with no significant changes in the morphology of its coma. The detailed analysis led to the following key findings:

a) Throughout the pre-perihelion observation period, 2I/Borisov exhibited a consistent appearance, behaving as an extended, diffuse cometary object with a homogeneous coma. No outbursts were detected during the follow-up.
b) The comet showed no signs of sudden brightness variations, indicating a steady sublimation process likely influenced by its prolonged exposure to interstellar space. Dust production remained relatively low, with Afρ typical values between 90 and 120 cm along the observational period. The observed moderate activity could be consequence of a surface covered by a refractory amorphous layer.
c) Our results obtained with mid-size class telescopes are comparable to those obtained with larger instruments. In consequence, there is lots of potential in these mid-size instruments to discover and track interstellar visitors. They can be as crucial for the discovery, as for the photometric follow-up of these elusive objects.
d) The overall weak and tenuous appearance of 2I/ Borisov also emphasizes the need of performing the searches of interstellar visitors using mid-size space telescopes. Unbiased observations from space are needed to increase the rate of discoveries of these elusive objects.
e) AI-powered algorithms can process vast amounts of astronomical data from ground-based and space-based telescopes, identifying faint or fast-moving objects that might otherwise be missed. Machine learning techniques can also classify cometary activity patterns, improving our understanding of sublimation processes and mass loss in interstellar comets.
f) The comet's magnitude shows a monotonical increase as consequence of its approach to perihelion, resulting in both a geometric brightening effect, and the gradual increase of sublimation processes due to volatile species.
g) Studying the activity and outgassing rates of interstellar comets provides information about the loss of mass of these bodies during close encounters to stars. If the loss rate is low they might preserve beneath the cosmic ray processed layer a pristine interior informing us on the bulk chemistry in their formative environments. In consequence, to implement a sample return capacity for a Comet Interceptor type mission, could be particularly relevant to gain insights into the nature of these interstellar visitors.
h) Our study exemplifies why preserving the quality of the night sky is crucial. Adopting dark-sky-friendly policies in urban planning and architecture plays a critical role in preserving observatory sites and ensuring optimal conditions for astronomical research. Enforcing outdoor lighting regulations, promoting full-cutoff fixtures, and implementing curfews for non-essential lighting can significantly mitigate light pollution, allowing astronomers to continue studying the cosmos with minimal interference. Enhanced image processing techniques can also help filter out artificial light interference, improving detection capabilities.

In view of the photometric results presented here, we can say that the first interstellar comet appearance occurred without sorrow or glory but providing valuable information on the cometary activity that we should expect of such a remarkable visitor. In any case, the discovery of 2I/Borisov is a clear confirmation that interstellar comets exist. Space agencies should coordinate to develop fast exploratory missions able to study the physical properties and nature of these elusive objects. Finally, given their highly eccentric orbits, these interstellar objects could exhibit very high relative velocities in their encounters with the Earth, so they could suddenly become challenging objects for life on Earth.

While light pollution remains a major obstacle in comet observation, telescopes with larger objectives enhance the range and visibility, allowing even faint comets to be detected under favorable sky conditions. In addition, satellite constellations affect our capacities to make long exposures for detecting these challenging and unexpected visitors. Establishing dedicated policies to protect the dark sky at the same time that we keep amateur astronomy in healthy condition to allow the discovery of new asteroids and comets could mark the difference among life and extinction.

Acknowledgements

We acknowledge support from the Spanish Ministry of Science and Innovation, research projects PGC2018-097374-B-I00 and PID2021-128062NB-I00 (PI: JMTR), funded by FEDER/Ministerio de Ciencia e Innovación – Agencia Estatal de Investigación, the Academy of Finland project no. 325806 "PlanetS" (PI: MG), and the Centre for Innovation and Transfer of Natural Sciences and Engineering Knowledge, University of Rzeszów, Poland (RPPK.01.03.00-18-001/10-00). The programme of development within Priority-2030 is acknowledged for supporting the research at UrFU. We thank Herbert Raab for his kind support with *Astrometrica* software. We also thank the feedback received from A. Fraser Gillan. TJO images at OAdM were scheduled by Toni Santana (UB).